\documentclass[]{aa}
\usepackage{graphics}
\begin{document}
\title{Large-scale shocks in the starburst galaxy NGC\,253}
  \subtitle{Interferometer mapping of a $\sim$600\,pc SiO/H$^{13}$CO$^+$ 
circumnuclear disk}
\author{S.Garc\'{\i}a-Burillo$^1$, J.Mart\'{\i}n-Pintado$^1$, A.Fuente$^1$, 
R.Neri$^2$}
%
%
\offprints {S.Garc\'{\i}a-Burillo}
\institute {Observatorio Astron\'omico Nacional (IGN), 
Apartado 1143, E-28800 Alcal\'a de Henares, Madrid, SPAIN (burillo@oan.es, 
martin@oan.es, fuente@oan.es)
\and
IRAM-Institut de Radio Astronomie Millim\'etrique, 300 Rue de la Piscine,
38406-St.Mt.d`H\`eres, FRANCE (neri@iram.fr)}

\date{Received , accepted }
%
%

\thesaurus{11.09.1 NGC\,253; 11.09.4; 11.11.1; 11.19.2}
\authorrunning{S.Garc\'{\i}a-Burillo et al}
\titlerunning{}
\maketitle

\begin{abstract}

This paper presents the first high-resolution SiO map made in an external galaxy. 
The nucleus of the nearby barred spiral NGC\,253  has been observed simultaneously in the v=0, J=2--1 line of SiO 
and in the J=1--0 line of H$^{13}$CO$^+$ with the IRAM interferometer, with a resolution of 7.5$\arcsec\times$2.6$\arcsec$.

Emission from SiO and H$^{13}$CO$^+$ is extended in the nucleus of NGC\,253.
The bulk of the SiO/H$^{13}$CO$^+$ emission arises from a (600pc$\times$250pc) circumnuclear disk (CND) 
with a double ringed structure. The inner ring, of radius r$\sim$60pc(4$\arcsec$), viewed edge-on 
along PA=51$^{\circ}$, hosts the nuclear starburst; the outer pseudo-ring opens out as a spiral-like 
arc up to r$\sim$300pc(20$\arcsec$). The kinematics of the gaseous disk, characterized by strong non-circular 
motions, is interpreted in terms of the resonant response of the gas to the barred potential. The inner ring would 
correspond to the inner Inner Lindblad Resonance(iILR), whereas the outer region is linked to the onset of a 
trailing spiral wave across the outer ILR(oILR). This scenario accounts for the unlike morphology of 
the maps observed in different molecules.   

Most notably, we report the detection of a molecular gas counterpart of the giant outflow of hot gas, previously seen 
in X-ray and optical lines, and tentatively identified as a dust chimney in the the 450$\mu$ continuum band. 
Two gas filaments appear in the SiO map to come out of the plane in the NGC\,253 nucleus at r$\pm$60pc, delimiting
the working surfaces of gas entrained by the hot wind.

The SiO shows a high average fractional abundance in the CND of $<$X(SiO)$>\sim$1.5$\times$10$^{-10}$. This is more than 
an order of magnitude above the predicted value of a PDR. Moreover, X(SiO) varies at least by an order 
of magnitude between the inner starburst region, which dominates the global emission, where we derive 
X(SiO)$\sim$1--2$\times$10$^{-10}$, and the outer region, where X(SiO) reaches a few 10$^{-9}$. 
SiO abundance is also significantly enhanced in the outflow (X(SiO)$\sim$3--5$\times$10$^{-10}$). 
Different mechanisms are explored to explain the unlike chemical processing of molecular gas within the nucleus. 
Large-scale shocks induced by the crowding of clouds orbits across the oILR in the outer region and 
by the outflow of hot gas seem to be by large the most efficient mechanisms in rising the Silicon abundance.

\end{abstract} 


\section{Introduction}

Among all tracers of the high density gas in spirals, the SiO molecule presents the most peculiar, and so far,
poorly understood chemistry. Galactic surveys of SiO clouds show that the relative abundance of this species 
varies by more than five orders of magnitude (the relative abundance, X(SiO), ranges 
from $<$10$^{-12}$ to 10$^{-7}$). SiO emission is practically
absent in quiescent dark clouds, indicating a high degree
of depletion in grains as X(SiO)$\sim$a few 10$^{-12}$ (Ziurys et al 1989). In contrast, SiO 
is enhanced in the dense and hot gas of GMCs forming massive stars, in particular, it appears in association with
bipolar outflows where  X(SiO)$\sim$10$^{-7}$-10$^{-8}$ (Mart\'{\i}n-Pintado et al 1992, Schilke et al 1997, Gueth et al
 1998). These high abundances have been interpreted as a signature of shock chemistry: SiO comes into the 
gas phase by the sputtering of grains by fast shocks. 
At significantly lower abundances (X(SiO)$\sim$10$^{-11}$-10$^{-10}$), SiO is predicted in 
Photon-Dominated-Regions (PDRs), complexes of molecular gas heated by UV photons produced by OB associations
(Janssen et al 1995, Walmsley et al 1999). Finally, SiO emission has been reported in molecular clouds
of the Galactic Center with X(SiO)$\sim$10$^{-8}$-10$^{-9}$ (Mart\'{\i}n-Pintado et al 1997, H{\"u}ttemeister et al 1998).
However these clouds show little evidence of recent star formation. 
The emission in these clouds might be connected with large-scale shocks rather than with star 
formation. Shocks are possibly driven by the bar potential or, more locally, by superbubbles and supernova remnants.

SiO emission was first detected in NGC\,253 by Mauersberger and Henkel 1991. A later study by 
Sage and Ziurys (1995) reported the detection of J=2--1 SiO emission in several starburst galaxies, including NGC\,253.
The authors concluded that the ratio I(SiO)/I(N$_2$H+) is an indicator of the SFR per unit mass. However, their conclusions on the 
origin of SiO emission are based on low resolution observations of only one transition.   
With the aim of studying the physical and the chemical parameters of SiO clouds,
we mapped in the J=5--4, 3--2 and 2--1 transitions of SiO the nuclei of three prototypical 
starbursts: M\,82, NGC\,253 and IC\,342 (Mart\'{\i}n-Pintado et al 1999, hereafter called {\bf MP99}).
First results of our survey, made with the 30m telescope, underlined the different influence of star formation 
in these galaxies. First, the inferred SiO abundances were seen to vary from galaxy to galaxy 
(from X(SiO)$\sim$10$^{-8}$ in IC\,342 to 
$\sim$10$^{-10}$ in M\,82), showing no correlation with changes on the SFR per unit mass. 
Physical conditions of SiO clouds, derived from the transition ratios, are also markedly different in our sample.
 
However, any further refinement in the interpretation of SiO emission required a better angular resolution.
In particular, it is crucial to elucidate which is the exact spatial extent of SiO emission in the nuclei of starburst 
galaxies. So far it is unclear if SiO emission is exclusively linked with the formation of massive stars, or 
alternatively, with the occurrence of large-scale shocks, as observations of clouds in our Galaxy seem to indicate.  
We want to address the question of which is the chemical scenario explaining the abundance of SiO in galaxies.

\begin{figure}[tphb]
\caption{{\bf (a,top)}: Emission contours of the 3mm (86.7GHz) continuum source at the 
center of NGC\,253. 
$x$ and $y$ are offsets (in arcsec) with respect to the dynamical center derived in this work, 
at $\alpha_{J2000}$=$00^h47^m33.18^s$, $\delta_{J2000}$=-$25^{\circ}17'17.2''$; $x$ and 
$y$ axes are parallel to the 
major and minor axes of the stellar bar, respectively ($x$ runs parallel to PA$_{bar}$=68$^{\circ}$). 
Contours are -1.5, 1.5, 2.5, 4, 5.5, 8 and 13 to 64mJybeam$^{-1}$ by steps 
of 9mJybeam$^{-1}$. 1$\sigma$-noise level is 0.44mJybeam$^{-1}$. 
{\bf (b,middle)}: SiO(v=0,J=2--1) integrated intensity contours towards the 
center of NGC\,253. 
Contours are -0.2, 0.2 to 2.2Jy.kms$^{-1}$beam$^{-1}$ by steps of 0.25Jy.kms$^{-1}$beam$^{-1}$.
1$\sigma$-noise level in the integrated intensity is 0.07Jy.kms$^{-1}$beam$^{-1}$. Orientation as in Figure 1a. 
{\bf (c,bottom)}: same as (b) but for the J=1--0 line 
of H$^{13}$CO$^+$, with same contours, noise level and orientation.}
\end{figure}

In this paper we study at high-resolution the SiO(v=0,J=2-1) emission in the nucleus of NGC\,253, 
a highly inclined barred Sc spiral representing the archetype of a nuclear starburst. 
The bulk of its large infrared luminosity L$_{IR}\sim$1.6$\times$10$^{10}$L$_{\odot}$
originates in the inner 300\,pc (Telesco and Harper 1980). The observed infrared spectrum of the nucleus (Telesco and 
Harper 1980; Engelbracht et al 1998) as well as the radio continuum observations made at centimeter wavelengths 
(Antonucci and Ulvestad 1988, Ulvestad and Antonucci 1991) support the starburst hypothesis.
There is also evidence of a giant outflow of hot gas powered by the starburst, that comes out of the galaxy plane
(Fabbiano and Trinchieri 1984, McCarthy et al 1987, Schulz and Wegner 1992).
The high molecular gas mass of the NGC\,253's nucleus ($\sim$2--3$\times$10$^{8}$M$_{\odot}$; Canzian et al 1988;
Mauersberger et al 1996) explains the high SFR. Water maser emission detected by Ho et al (1987) reveals strong star 
formation activity. The gas reservoir might have been driven inward by the barred potential 
first identified by Scoville et al 1985 in a K-band image of the galaxy. 
Several interferometer maps showing the molecular gas distribution of NGC\,253's nucleus have been published 
(CO: Canzian et al 1988; HCN: Paglione et al 1995; CN: H{\"u}ttemeister and Aalto 1998, HCO$^{+}$: Carlstrom et al 1990; 
OH: Turner 1985; H$_2$CO: Baan et al 1997; CS: 
Peng et al 1996, hereafter {\bf P96})
However, no high-resolution image of the dense gas distribution (n(H$_2$)$>$10$^{5}$cm$^{-3}$) has been obtained so far, using
an optically thin tracer.    
To fill this gap, we have made simultaneous observations in the J=1--0 line of H$^{13}$CO$^+$,
a tracer of dense gas with low opacity.
Furthermore all the results for SiO and H$^{13}$CO$^+$ are discussed in parallel throughout the paper. In particular 
we study how the ratio of integrated intensities I(SiO)/I(H$^{13}$CO$^+$) changes 
across the nucleus and discuss how this result is used to get the variation of the absolute abundance of SiO.   

\section{Observations}

Observations of NGC\,253 were made with the IRAM array at
Plateau de Bure (France) between June and August 1998. We simultaneously observed
the (v=0,J=2--1) line of SiO (at 86.847\,GHz) and the (J=1--0) line of 
H$^{13}$CO$^+$ (at 86.754\,GHz) using the standard BC set of 4-antenna 
configurations. The 55$''$ primary beam field of the interferometer was phase-centered 
at $\alpha_{J2000}$=$00^h47^m33.0^s$ and 
$\delta_{J2000}$=-$25^{\circ}17'18.4''$.

The SIS receivers were tuned for single side band operation at 86.731\,GHz yielding 
system temperatures of $100-200$\,K. The spectral correlator was adjusted 
to give a contiguous bandwidth of 420\,MHz. This is equivalent to a velocity range of 
1500\,kms$^{-1}$; the correlator was centered at 86.731\,GHz in order to cover both 
transitions. The effective frequency resolution was set to 2.5\,MHz, or 
equivalently 8.64\,kms$^{-1}$ at this frequency. 

Visibilities were obtained using on-source integration times of 20 minutes 
interspersed with 4 minutes of instrumental amplitude and phase calibrations. These were made by observing  
0135-247 and 2345-167. The atmospheric phase noise on the 
most extended baselines ranged from 20$^{\circ}$ to 40$^{\circ}$, consistent
with a seeing of 0.8--1.2'', typical for summer weather conditions. 
The absolute flux density scale was established on the basis of 
cross-correlations on the radio star MWC349 (with $S_{87GHz}=950$\,mJy). This is in 
full agreement with the measured interferometer efficiency and should 
be accurate to 10\%. The receiver passband shape was determined on 3C454.3 
and its accuracy is better than 5\% throughout the observing run.

Cleaned maps were obtained from the visibilities using the standard IRAM 
package. 
The maps were $256\times 256$ pixels in extent, with a pixel 
size of $0.4''$. The synthesized beam, as determined by fitting a Gaussian 
to the dirty beam, is $7.5''\times 2.6''$, oriented 
north-south (PA=180$^{\circ}$). Due to the low declination of the galaxy, the uv 
plane 
is unequally sampled with a maximum spacing of 70\,m in the north-south 
direction and of 250\,m in the east-west. The corresponding east-west linear 
scale at the distance of the source, assuming $D=3.4$\,Mpc (Sandage and Tammann 1975),
 is 42\,pc. A continuum map was obtained by averaging visibilities over a bandwidth 
chosen to be free of line emission. In practice we used spectral channels with 
velocities 200\,kms$^{-1}$ above and $-200$\,kms$^{-1}$ below the galaxy's 
systemic velocity. For the effective 235\,MHz bandwidth used to obtain continuum data, we 
derived a one $\sigma$ point source sensitivity limit of 0.44\,mJy/beam. 
The rms noise level in 5\,MHz wide channel maps, as determined from the 
dirty maps after subtraction of the continuum, is 2\,mJy/beam. This  
corresponds to an rms brightness temperature of 17 mK for the synthesized 
beam size and is consistent with a total on-source integration time of 
15 hours and a mean system temperature of 150\,K.

We have estimated the zero-spacing flux lacking in our maps. 
Taking into account that the shortest spacing measured by the interferometer is 
$\sim$20m,
we expect to filter out scales $\sim$30$\arcsec$. Compared to the 30m data of 
{\bf MP99}, only $\sim$20$\%$ of the single-dish flux in both SiO and H$^{13}$CO$^+$ is 
missing within the primary beam. Therefore, our data offer an unbiased picture of the total 
SiO and H$^{13}$CO$^+$ gas content in the nucleus of NGC\,253.

We assume that the orientation of the NGC\,253 disk is defined by the angle 
PA$_{disk}$=51$^{\circ}$, and inclination i=78.5$^{\circ}$. The orientation of the major axis of the stellar bar, 
along PA$_{bar}$=68$^{\circ}$, is used on purpose in some figures throughout the paper.

\section{The interferometer maps}

\subsection{The radio continuum emission}

The 3mm continuum emission contours are plotted in Figure 1a.
Despite its compactness, this continuum source, with a peak flux of 70mJy/beam, 
is resolved by our beam, with 80$\%$ of the flux coming from the central source with a deconvolved size 
of $\sim$8$\arcsec\times$4.3$\arcsec$, oriented along the disk major axis 
(PA$\sim$50$^{\circ}$). 
The integrated flux amounts to 0.250$\pm$0.008Jy; extrapolating fluxes at 6cm and 2cm(Turner and Ho 1983), 
we estimate that nonthermal and thermal free-free emission account each for one third of 
the observed 3mm continuum flux. 
The remaining  third would come from emission by dust.  
Although our value for the integrated flux agrees within
the errors with that measured at 85GHz by Carlstrom et al 1990 (0.300$\pm$0.050Jy), it is slightly lower 
than that derived by {\bf P96} at 98GHz (0.320$\pm$0.030Jy).

The better sensitivity of our maps, compared with prior observations (Carlstrom et al 1990; {\bf P96}) 
reveals the presence of two previously undetected sources located along the bar major axis 
at offsets x=-10$\arcsec$ and x=10$\arcsec$. These new sources appear marginally as extensions in the
1.3mm maps of Kr{\"u}gel et al 1990. This suggests that they correspond mostly to dust emission. However
the nature of these 3mm new sources is unknown. Comparison of our data 
with measurements at other wavelengths is not straightforward because of the missing flux in the images
at the centimeter range and, on the other hand, owing to the lack of resolution of the existent maps 
at 1.3mm, 0.8mm and 0.45mm. Before high angular resolution observations at short millimeter and 
submillimeter wavelengths are obtained, little can be said on the nature of these new sources.     
   
\begin{figure*}[tphb]
\caption{SiO(v=0,J=2-1) velocity-channel maps observed with the 
IRAM interferometer of Plateau de Bure with a resolution (HPBW) of 7.5$\arcsec\times$2.6$\arcsec$ 
(PA=180$^{\circ}$). 
Absolute coordinates are 2000.0 (the derived dynamical center is indicated by 
the cross at $\alpha_{J2000}$=$00^h47^m33.18^s$, $\delta_{J2000}$=-$25^{\circ}17'17.2''$).
Velocity-channels range from v=361.5kms$^{-1}$ to v=93.8kms$^{-1}$ by steps of 
-8.6kms$^{-1}$.
Contour levels are -6, 6, 8, 10, 12, 15, 19 and 23mJybeam$^{-1}$. 1$\sigma$-noise level
is 2mJybeam$^{-1}$. Axes are ($\alpha$,$\delta$).}
\end{figure*}

\begin{figure*}[tphb]
\caption{Same as Figure 2 but for the J=1--0 transition of H$^{13}$CO$^+$ with the same 
levels, velocity spacing and axes orientation.}
\end{figure*}

Contrary to {\bf P96}, we see no extension of continuum emission to the 
southwest of the nucleus, interpreted as a signature of the giant outflow detected in X-rays.
Continuum emission of the dust outflow has only been tentatively detected at 0.45mm by 
Alton et al 1999, but it is absent in the 1.3mm and 0.8mm maps. 
We have extrapolated the flux measured at 0.45mm to 3mm and corrected for different resolutions,
assuming an emissivity law of $\sim\nu^2$. We predict 
a 3mm counterpart of 0.1mJy/beam for the dust outflow in NGC\,253, well beyond the noise level in our 
map (0.44mJy/beam) and a factor of $\sim$200 weaker than the 3mm extensions of {\bf P96}, that reveal
as spurious.

\subsection{The SiO and H$^{13}$CO$^+$ emission}     
     
Figure 2 shows the velocity-channel maps obtained 
in the (v=0,J=2--1) line of SiO at the centre of NGC\,253; figure 3 shows the same but in 
the (J=1--0) line of H$^{13}$CO$^+$. 
Gas emission is unevenly distributed in the velocity range 
[120-353kms$^{-1}$]. This arises from a circumnuclear disk (CND) extending over 
600\,pc$\times$150\,pc. The disk, seen at high inclination (i=78.5$^{\circ}$), is radially resolved and 
displays an overall rotating pattern. 
 
Although the dominant trend is circular rotation, we also find emission at highly non-circular velocities 
in both tracers. The emission is patchy in 
both lines, most noticeably, the SiO and H$^{13}$CO$^+$ clumps show no one-to-one correspondence 
over the map: as discussed below (section 7), the 
T$_{mb}$(SiO)/T$_{mb}$(H$^{13}$CO$^+$) ratio has a wide range of variation in the nucleus of NGC\,253 
(i.e. T$_{mb}$(SiO)/T$_{mb}$(H$^{13}$CO$^+)\sim$0.5--8)).

The CND also shows structure along the apparent minor axis.
The velocity-integrated intensity maps of SiO (I$_{SiO}$) and H$^{13}$CO$^+$ 
(I$_{H^{13}CO^+}$), shown in Figures 1b-c, illustrate the CND substructure.
Spatial coordinates are arcsecond offsets along the principal axes 
of the near infrared (NIR) stellar bar (Scoville et al 1985).
The bar major axis (denoted x) is at PA=68$^{\circ}$, with 
x$>$0 eastwards (the bar minor axis, denoted y, lies along PA=158$^{\circ}$, with 
y$>$0 northwards). Figure 1b shows that the SiO emission mostly comes from two-nested rings 
centered at (0,0) and roughly oriented parallel to the x axis. A similar morphological description 
applies to H$^{13}$CO$^+$ map (Figure 1c), although here the inner ring orientation 
deviates significantly from the bar major axis.

The inner ring ({\bf I}), of average diameter D$_{I}\sim$8$\arcsec$(120pc), 
contains 60-70$\%$ of the total flux in the two lines and it corresponds to the region of the 
massive starburst in NGC\,253 seen in different tracers. 
The emission of the outer ring ({\bf II}), of average diameter D$_{II}\sim$24$\arcsec$(360pc), 
contains 20-30$\%$of the total flux. The two nested rings share as common center the peak of the 3mm continuum 
source. Emission in {\bf I} is detected within a large velocity range for 
the two edges of the ring ($\sim$120kms$^{-1}$). This spread is symmetrical with 
respect to the continuum source locus  (a similar symmetry is seen to hold for {\bf II}). 
We therefore identify it as the dynamical center of NGC\,253 ($\alpha_{J2000}$=$00^h47^m33.18^s$ 
and $\delta_{J2000}$=-$25^{\circ}17'17.2''$). This position coincides, within the errors, 
with the dynamical center determined using optical (Watson et al 1996), infrared (Sams et al 1994, Boker et al 1998) 
and radio data (Turner and Ho, 1985).

Although most of the emission arises from the double ringed source ({\bf I-II}), we have also
detected gas outside the x axis: the CND structure along the minor 
axis is partly resolved. Although part of the emission can be accounted by the gas distribution in the CND disk
seen in projection (NGC\,253 is not fully edge-on), at places, the emission lies far away from the NIR bar major axis, and
most noticeably, this corresponds to gas clouds showing highly non-circular motions and a high relative abundance of SiO 
(T$_{mb}$(SiO)/T$_{mb}$(H$^{13}$CO$^+$)=3-5)). In particular, the SiO emission channels of 
Figure 2 in the range [284-335kms$^{-1}$] show receding gas (v-v$_{sys}$=80kms$^{-1}$; where 
v$_{sys}$=235kms$^{-1}$) in the SE quadrant, where {\it it should be} approaching.  
As will be discussed below, the origin of this {\it anomalous} component is related with the giant outflow detected 
in X-ray and optical emission lines. This does not have the same origin as {\bf I-II}.  
However, there is a H$^{13}$CO$^{+}$ clump at (x,y)$\sim$(1,8) which shows no SiO counterpart. Contrary to the 
anomalous component, this might correspond to disk gas seen in projection. This clump has a low 
ratio T$_{mb}$(SiO)/T$_{mb}$(H$^{13}$CO$^+$)$\sim$0.3, and most important, it lies close to the minor axis 
of the galaxy appearing at velocities close to v$_{sys}$.

\section{The CND structure}

To study the CND structure we have obtained the peak brightness 
flux distributions in the two lines (B$^{peak}$(x,y)=max[T$_{mb}$]$_v$(x,y)). This allows us to filter out 
the lower spatial frequencies enhancing the most intense features in the maps.
Then we deprojected B$^{peak}$ onto the galaxy plane, in order to get a sharp picture of the 
CND, viewed face-on. The result is shown in Figures 
4a-b. The peaks of emission of ring {\bf I} are rather aligned 
with the major axis of the disk, tilted 20$^{\circ}$ clockwise 
with respect to the bar major axis. The tilt is more pronounced for H$^{13}$CO$^+$ (Figure 4a). 
On the contrary, the peaks of emission of ring {\bf II} open out, go across the line of the 
bar major axis and delineate a spiral-like arc. 
The face-on picture of the CND of Figure 4a consists of a gaseous spiral-like ridge ({\bf II}) ending up in an 
unresolved ring {\bf I} viewed edge-on and aligned with the major axis of the disk. 

\begin{figure}[tphb]
\caption{{\bf (a)}: H$^{13}$CO$^+$(1--0) peak-brightness flux contours
in the CND of NGC\,253, deprojected onto the galaxy plane ((X$_G$,Y$_G$); where X$_G$ runs parallel
to the major axis of the disk along PA$_{disk}$=51$^{\circ}$), displayed from
7 to 19mJy/beam by steps of 2mJy/beam. The inner region is a nearly edge-on ring encircling the starburst, whereas
the outer region opens up towards larger PAs, delineating a spiral-like arc.
{\bf (b)}: SiO$(v=0,J=2-1)$ peak-brightness temperature contours 
in the CND of the galaxy, as seen in the plane of the sky in (x,y) coordinates as defined in text. 
Same contours as (a). The thick line high-lights the 
orientation of the N and S filaments.}

\end{figure}

Interferometer observations by Canzian et al (1988) have shown that the bulk of molecular gas in the CND, 
traced by the 1--0 line of $^{12}$CO, is distributed in a bar-like 
source, oriented in the direction of the NIR bar. 
The recent $^{13}$CO map of H{\"u}ttemeister and Aalto (1998) confirms this view. 
In contrast, the distribution of dense gas (n(H$_2$)$>$10$^{5-6}$cm$^{-3}$) is similar to 
the B$^{peak}$ maps of SiO and H$^{13}$CO$^+$. 
The CN(1--0) and HNC(1--0) maps of H{\"u}ttemeister and Aalto (1998) show a centrally condensed inner 
disk of diameter $\sim$8$\arcsec$, oriented along PA=45$^{\circ}$-50$^{\circ}$, while the outer disk 
tilts farther towards PA=70$^{\circ}$, delineating a spiral-like arc that goes 
across the major axis of the stellar bar. The same applies for the HCN(1--0) map of Paglione et al 1995: the 
inner disk is better aligned with the disk major axis PA=45$^{\circ}$, whereas the outer 
ring tilts progressively towards PA=70$^{\circ}$.

There is little doubt that NGC\,253 contains a stellar bar oriented along 
PA=68$^{\circ}$ and extended 300$\arcsec$ on the sky, which
sets a lower limit for its corotation radius at R$_{COR}$=150$\arcsec$ (Scoville et al 1985).
{\bf P96} first explained the distribution of the 
CS(2--1) emission in terms of a bar potential. However the parameters of the bar were chosen arbitrarily: 
the bar corotation is placed at 40$\arcsec$, a factor 
of $\sim$4 too short, compared with the estimated bar semi-major axis ($\sim$150$\arcsec$). 
The NIR image of the bar is distorted by dust extinction. There are two dust lane ridges 
offset with respect to the bar major axis. Assuming that the NW is the near side of 
NGC\,253 (De Vaucouleurs 1958), the two offset dust lanes 
inside the bar represent a trailing spiral wave. This pattern is an indication of the 
existence of two Inner Lindblad Resonances (outer=oILR and inner=iILR; see  Athanassoula (1992)).
The morphology of the gas response in the presence of two ILRs has been analysed by several authors (see Buta and Combes 1996
and references therein). Owing to dissipation by cloud-cloud collisions, 
the gas cannot follow the periodic orbits of the stars in a bar (x$_1$=parallel, between corotation and
oILR or x$_2$: antiparallel, between the oILR and the iILR). The gas response always leads 
the stellar orbits, giving a trailing spiral inside the bar corotation that goes across the oILR. 
It is tempting to identify the spiral-like {\bf II} as the oILR. The gas compression along the spiral ridges should be 
best traced, as observed, by the higher density tracers of molecular gas. In contrast, the nuclear spiral across 
the oILR is washed out in the $^{13}$CO and $^{12}$CO maps tracing the low or moderately dense molecular 
gas (n(H$_2$)=10$^{3-4}$cm$^{-3}$).

Moreover, the proposed resonance loci of the bar naturally explain the double-ringed appearance of NGC\,253 (Figures 1bc), and 
how the nuclear starburst might have been onset. 
When a barred galaxy has two ILRs, as NGC\,253, gas tends to 
accumulate at the iILR, as a consequence of secular evolution. The engine behind the radial inflow of gas is 
double. Gravity and viscous torques cooperate in bringing the gas closer to the nucleus. 
Gravity torques act on the gaseous spiral expected to form between the iILR and oILR. The spiral
is a superposition of leading and trailing waves. The leading wave dominates at the crossing of the 
iILR, whereas the trailing wave develops at the crossing of the oILR (Buta and Combes 1996). 
The gravity torques change their sign at the crossing of the resonances, and also when we move from the trailing 
to the leading spiral (Combes 1994a, 1994b). The net result would be the formation 
of two pseudo-rings at the iILR and the oILR, with similar time-scales. However, viscous torques being always negative, may break 
this symmetry making the gas migrate towards the iILR in a few dynamical times ($\sim$10$^{7-8}$years) 
and therefore progressively depopulate the oILR. The oILR may end up as a trailing spiral-like pseudo-ring 
(ring {\bf II} in NGC\,253), whereas the leading spiral vanishes into the inner ring (similar to ring {\bf I}).

It seems that the morphology of the SiO/H$^{13}$CO$^+$ maps and how they compare to other molecular gas tracers can be explained 
by gas accumulation at the ILRs of the bar in NGC\,253. 
Arnaboldi et al 1995, used the H$\alpha$ major axis kinematics to make an independent estimation of the bar resonances loci. 
Within the epicyclic approximation,
they identified two ILRs in the nucleus of NGC\,253. 
Assuming a lower limit for the bar corotation, R$_{cor}$=3.9kpc, they derived an oILR at 300pc (20$\arcsec$) and 
the iILR near the center, in agreement with what is proposed in this work. 
Studying the gas kinematics in the SiO/H$^{13}$CO$^+$ CND will allow to test the location of the principal 
resonances and understand the relation between the starburst and the bar.

\section{Kinematics of the CND}     

There is a wealth of published work on the dynamics of the ISM in the nucleus of NGC\,253.
The determination of a rotation curve (v$_{rot}$) for the nucleus of NGC\,253 has been very 
controversial because of extinction (Ulrich 1978). 
The high-sensitivity H$\alpha$ data of 
Arnaboldi et al 1995, already showed a velocity gradient more symmetrical and steeper than seen in previous
optical surveys.

Figures 5a-b show the SiO and H$^{13}$CO$^+$ position-velocity diagrams taken along the kinematical 
major axis of the disk of NGC\,253  (x' axis). In view of the high inclination of NGC\,253, the 
terminal velocity method would be well suited to derive v$_{rot}$ from the major axis kinematics. 
However the patchiness of the SiO/H$^{13}$CO$^+$ emission makes any fit uncertain. 
There are strong oscillations of the terminal velocities as a function of radius, revealing an uneven 
gas distribution and/or the presence of non-circular motions. 
As a zero-order approach we will adopt v$_{rot}$ derived by Canzian et al(1988) from CO. 
This species is the best tracer of molecular gas at a 
large-scale, from densities as low as n(H$_{2}$)$\sim$10$^{3}$cm$^{-3}$, and although non-circular motions
should also affect the observed kinematics, the estimated v$_{rot}$ would be closer to the axisymmetric value. 

\begin{figure*}[tphb]
\caption{The major axis (along PA=51$^{\circ}$) position-velocity diagrams taken at the center of 
NGC\,253 in the SiO(v=0,J=2-1) 
{\bf (a,left)} and H$^{13}$CO$^+$ {\bf (b, right)} lines. x' offsets along major 
axis (PA=51$^{\circ}$)
are referred to the dynamical center derived in this work: $\alpha_{J2000}$=$00^h47^m33.18^s$ 
and $\delta_{J2000}$=-$25^{\circ}17'17.2''$. v$_{sys}$(LSR) is taken at 235kms$^{-1}$. Contours are -0.004, 0.004, 
0.006, 0.008, 0.010, 0.012, 0.015, 0.019 and 0.023Jybeam$^{-1}$. 
Our effective resolution along the major axis is 
$\Delta$x'$\times\Delta$v=3.2$\arcsec\times$8.6kms$^{-1}$. The rotation curve derived from CO (v$_{rot}$(CO)) is superposed
to Fig 5a. The romboid-like parallelogram delimits the p-v space occupied by bar-driven orbits inside the ILR domain
(see text).}    
\end{figure*}

The centroid of SiO/H$^{13}$CO$^+$ emission in the major axis p-v plots follows closely v$_{rot}$(CO), but most noticeably, 
the emission shows a large velocity spread around ($\Delta_v$=100-120kms$^{-1}$; see Figures 5a-b). 
The observed linewidths could be partly due to the smearing of v$_{rot}$ within our beam. 
Assuming i=78.5$^{\circ}$, we expect to get emission from clouds at y$_{depr}$=$\pm$9-10$\arcsec$ 
(y$_{depr}$ is the deprojected distance from which emission is seen by our beam along y'). 
Along the major axis we get emission from x'=$\pm$1.6$\arcsec$. 
v$_{rot}$(CO) follows a rigid body law with a velocity gradient of 6kms$^{-1}$/$\arcsec$, 
from which we derive an upper limit for the total velocity broadening  of $\sim$60kms$^{-1}$ 
within our beam, i.e., a factor of $\sim$2 lower than the observed velocity spread. 

The existence of a large velocity dispersion in the CND gas cannot be excluded (we would need 
$\sigma_v\sim$40-50kms$^{-1}$ to explain the observed linewidths). However, an explanation of 
the SiO/H$^{13}$CO$^+$ linewidths in terms of unresolved non-circular motions is more plausible
as it is supported by additional observational evidence.

Firstly, a part of the emission is detected at velocities that cannot be accounted by {\it any} 
circular rotation law. These regions in the major-axis p--v plots correspond 
to the (x',v)-quadrants (x'$>$0,v$>$v$_{sys}$) and (x'$<$0,v$<$v$_{sys}$)). 
SiO/H$^{13}$CO$^+$ gas emission fills unevenly a romboid-like parallelogram in Figures 5a-b. 
The symmetry of the above pattern is remarkable and it reminds of the signature of bar-driven elliptical orbits 
developed inside the ILR domain (Binney et al 1991; Garc\'{\i}a-Burillo and Gu\'elin 1995, Fux 1998, 
Garc\'{\i}a-Burillo et al 1999). 
The parallelogram boundary would correspond to the inner non-selfintersecting x$_1$ orbit 
(called the {\it cusped} orbit). Inside, gas clouds lie along precessing x$_2$ orbits whose envelope delineate a 
spiral across the oILR; these orbits are highly elliptical and produce the observed non-circular motions.   
The velocity gradient is reversed for a SiO component in the 
major axis p-v plot between x'=-4$\arcsec$ and x'=4$\arcsec$. The latter was also reported by {\bf P96} in the CS
p-v plot and and by Ananthamaraiah and Goss(1996) using recombination-line data. {\it Apparent} counter-rotation 
can also be explained by bar orbits (see Figure 8 of  Garc\'{\i}a-Burillo and Gu\'elin 1995).     
The SiO minor axis position-velocity diagram, displayed in Figure 6a, suggests also that gas distribution 
in the nucleus deviates from axisymmetry. The cut along the minor axis is markedly asymmetrical: 
the centroid of redshifted gas lies 1$\arcsec$-2$\arcsec$ above the major axis locus, whereas blue-shifted 
gas lies 1$\arcsec$-2$\arcsec$ below.

\begin{figure*}[tpbh]
 \caption{{\bf a} SiO strip along the minor axis of the galaxy (PA=141$^{\circ}$) with 
same contours as Figure 5. The effective resolution is
$\Delta$y'$\times\Delta$v=4$\arcsec\times$8.6kms$^{-1}$. We also display 
in {\bf b} the SiO position-velocity diagram parallel to the minor axis, but taken at 
x=4$\arcsec$, showing the presence of gas emission at highly non-circular 
motions (inside the high-lighted region).}
\end{figure*}

Ananthamaraiah and Goss(1996) reported the existence of a complex kinematic subsystem in the inner 150pc of NGC\,253, 
using recombination-line data. The nucleus would host three gaseous disks; among the anomalous components, one exhibits 
rotation in a plane perpendicular to the galactic disk, and the inner disk is {\it apparently} counter-rotating.
These components are introduced to explain deviations from circular motions similar to what
is shown in our data.  On the basis of this somewhat {\it ad hoc} decomposition, the authors find evidence of 
a past merger involving a small mass companion. However, the merger hypothesis is unable to explain the undisturbed morphology 
and kinematics of the outer disk (Olson and Kwan, 1990). 
The distorted kinematics of the NGC\,253 nucleus, including the existence of strong non-circular motions
detected also in the recombination-line data, are more readily explained in terms of a bar-driven gas flow.

\section{A molecular gas counterpart of the NGC\,253 nuclear outflow}

In the previous section we have argued that the bulk of the SiO/H$^{13}$CO$^+$ emission in the CND fits within 
the bar-driven gas flow scenario. 
However, we reported in section 3 the detection of an anomalous component which might be far from the CND plane. 
The off-axis gas, mostly detected in the SiO line (T$_{mb}$(SiO)/T$_{mb}$(H$^{13}$CO$^+$=3-5), is best seen towards the south 
({\it down the plane} of the CND). The southern SiO plume starts at the base of the starburst ring {\bf I} and it 
reaches a {\it height} of $\delta$y=10$\arcsec$ (measured from the major axis of the bar; see Figure 4b). The gas shows also strong 
deviations from circular motion ($\Delta_v\sim$100-150kms$^{-1}$) and its emission spreads over $\sim$200kms$^{-1}$
(see Figure 6b).  
There is a much weaker and point-like counterpart of the southern SiO-plume towards the North.
 The northern gas shows also strong noncircular motions 
($\Delta_v\sim$-150kms$^{-1}$) but it is here characterized by a narrow  profile ($\sim$17kms$^{-1}$).
SiO spectra towards these positions are displayed in Figure 7.
The Northern and Southern plumes (hereafter called N and S) can be connected by a line  
that goes across the dynamical center; the N-S line has a PA=131$^{\circ}$ measured from North (see Figure 4b).         
This orientation places the anomalous component close to the galaxy minor axis (PA=141$^{\circ}$) where no contribution 
from rotation curve in the radial velocity gradient is expected.   
Moreover it is hard to explain non-circular motions of $\Delta_v\sim$100-150kms$^{-1}$ due to bar forcing in the plane 
of the galaxy. 
\begin{figure*}[tphb]
 \caption{We compare the SiO(v=0,J=2-1) (thick line) and H$^{13}$CO$^+$ (thin line) spectra at several 
positions: (x,y)=(0$\arcsec$,0$\arcsec$){\bf (left bottom)} in the dynamical center, 
(x,y)=(3.2$\arcsec$,3.2$\arcsec$){\bf (left middle)} in region {\bf I}, 
(x,y)=(-17.6$\arcsec$,1.6$\arcsec$){\bf (left top)} in region {\bf II}, 
(x,y)=(3$\arcsec$,-8$\arcsec$){\bf (right top)} in the southern outflow and 
(x,y)=(-5$\arcsec$,8$\arcsec$){\bf (right bottom)} in the northern outflow; the spectra
illustrate the large variations of the T$_{mb}$(SiO)/T$_{mb}$(H$^{13}$CO$^+$) ratio 
in the CND. Y axis scale is in Jy/beam.}    
\end{figure*}

There is compelling evidence for the existence of a giant outflow of gas in the nucleus of NGC\,253.  
Demoulin and Burbidge(1970) and Ulrich (1978) were the first to report the presence of an ionized gas outflow, 
based on the analysis of gas kinematics derived from H$\alpha$ spectroscopy. Soft X-ray emission from the 
giant outflow was later discovered by Fabbiano and Trinchieri (1984). The X-ray nebula is more extended 
to the southern part of the galaxy minor axis and was interpreted as thermal emission of gas heated by fast shocks,
driven by a bipolar wind coming from the starburst. Further support for this scenario is provided by 
the distribution of ionized gas in the outflow, which adopts the form of filaments that border the X-ray nebula 
(McCarthy et al 1987). Furthermore, Schulz and Wegner (1992) detected line-splitting in NII, SII and HII, 
consistent with the ionized gas lying across the conical working surface of the X-ray nebula. The gas entrainment 
by the hot wind may cause shock-heating and subsequent emission. The observed line-splitting, together with the 
inferred expansion velocities ($\sim$390kms$^{-1}$), are the main arguments supporting the shock-heating scenario.

Very recently, Alton et al 1999 have reported the tentative detection of submillimeter emission at 450$\mu$ coming from
one filament associated with the S outflow. 
The location of the S filament agrees reasonably well with
the one detected in SiO/H$^{13}$CO$^+$ (see above and Figure 4b). The dust mass contained in the filaments, which partly depends
on the dust temperature, cannot be easily estimated from the detection at just one wavelength.
No counterpart of the dust outflow is seen neither at 1mm nor at 3mm, as expected from the flux detected at 450$\mu$
(see discussion of section 3.1; see also Kr{\"u}gel et al, 1990). However we can try to estimate the dust column density towards the S 
filament from the flux reported by Alton et al 1999, corrected to our beam. We assume a dust emissivity spectral 
index of $\beta$=2, the emissivity law is taken from Chini et al (1997),
 and adopt a dust temperature in the range T$_{dust}$=13K-37K (similar to
the model that fits the M82 dust outflow: Alton et al 1999). 
We calculate a range of N$_{dust}\sim$0.6--4$\times$10$^{4}$M$_{\odot}$ towards the S filament, within our beam. 
If M$_{gas}$/M$_{dust}$=100, this results implies 
N(H$_2$)$\sim$0.6--4$\times$10$^{6}$M$_{\odot}$/beam.  
As discussed in section 7, we can estimate N(H$_2$) in the S filament where H$^{13}$CO$^+$ is detected, 
using N(H$^{13}$CO$^+$) derived from a Large Velocity Gradient (LVG) transfer model and assuming 
X(H$^{13}$CO$^+$)=10$^{-10}$ (see discussion of section 7 for details). We calculate 
N(H$_2$)$\sim$3.5$\times$10$^{6}$M$_{\odot}$/beam (4.7$\times$10$^{22}$cm$^{-2}$), 
roughly in agreement with the value estimated from the dust emissivity. Therefore no
contradiction exists between the failed detection of the S outflow in the continuum and the successful detection of the outflow 
in the two lines of SiO and H$^{13}$CO$^+$.

If the molecular CND is locally disrupted and it is entrained by the hot wind out of the plane, we can expect that the
geometries for the working surfaces of molecular and ionized gas should be alike. The parameters of Schulz and
Wegner's model are an expansion velocity of 390kms$^{-1}$ for the conical outflow coming out of the plane, 
and a half-opening angle of 25$^{\circ}$. Projection effects along the line of sight would make appear two velocities
at each position of the outflow (N or S). The emission 
from the S(N) filament should appear either at v$_{rad}$=-10kms$^{-1}$ (120kms$^{-1}$) or at 
v$_{rad}$=300kms$^{-1}$ (440kms$^{-1}$). The velocities observed in the SiO lines for the N 
(v$_{rad}$=100kms$^{-1}$) and S filaments (300kms$^{-1}$) confirm the predictions of Schulz and 
Wegner's model, originally intended to fit the observed line-splitting in the optical lines.    
Turner (1985) was the first to report the existence of a molecular gas ejection event in the NGC\,253 nucleus. 
However, the location of the OH plume hardly fits the morphology and kinematics of the ionized or molecular outflow:  
the OH gas is ejected from the nucleus towards the NE quadrant (along PA=6$^{\circ}$) and it reaches $\sim$1kpc height (see Fig 11
of Turner(1985) for details).

We conclude that the distribution and kinematics of the off-axis component detected in our maps
are easily accounted for if we suppose that the emission comes from molecular gas in an outflow 
leaving the nucleus, out of the galaxy plane. Furthermore, an estimate of the molecular gas mass involved in 
the outflow seems to be consistent with the dust emissivity of the filaments.

\section{The T$_{mb}$(SiO)/T$_{mb}$(H$^{13}$CO$^+$) and  N(SiO)/N(H$^{13}$CO$^+$) ratios}

The line temperature ratio R$_{Tmb}$=T$_{mb}$(SiO)/T$_{mb}$(H$^{13}$CO$^+$) 
shows large variations (0.5--8) in the nucleus 
of NGC\,253; these variations however show a systematic trend. For the sake of simplicity, we first examine the ratio 
of velocity-integrated emission R$_{I}$=I(SiO)/I(H$^{13}$CO$^+$) averaged across those regions where R$_{I}$ shows a 
characteristic value. We have chosen as cases of study the dynamical center, regions {\bf I-II} and the
outflow components (N--S). Table 1 shows the mean values of R$_{I}$ and its range of variation; Figure 7 
shows representative spectra for the regions studied. We see a clear trend in the ratios, which increase as one 
moves away from the vicinity of the starburst. The average ratios are 
R$_{I}\sim$1.0, for the center, R$_{I}\sim$1.1 for the starburst ring {\bf I},
R$_{I}\sim$3 for the outer region {\bf II} and R$_{I}\sim$2 for the outflow. 
We will analyse the range of variation of R$_{Tmb}$ for each region defined above, and derive the column density 
ratios N(SiO)/N(H$^{13}$CO$^+$) via a Large Velocity 
Gradient (LVG) scheme. Our objective is to study the range of physical parameters and chemical abundances of 
SiO clouds in the CND. 


LVG solutions depend on three parameters: 
the density (n(H$_2$)), the kinetic temperature (T$_K$) and the abundance of the species. 
The kinetic temperature of molecular gas in the center of NGC\,253 has been derived by several authors using 
multitransition studies in $^{12}$CO and its isotopes ($^{13}$CO and C$^{18}$O), CS and HCN; all indicate an 
average value of T$_{k}>$50K. We have adopted T$_{k}$=50K as representative for the SiO clouds
and taken n(H$_2$) from the fit to the ratios R$_{32}$=I(SiO(3--2))/I(SiO(2--1)) and R$_{54}$=I(SiO(5--4))/I(SiO(3--2)) derived 
by {\bf MP99}, using single-dish 30m spectra. {\bf MP99} found little evidence of a 
density decrease for the SiO clouds, at least within the inner r$\sim$20$\arcsec$ of the nucleus.
They derived an average n(H$_2$)=5$\times$10$^{5}$  assuming a common filling factor for the three lines.
We will take this estimation as representative for SiO and H$^{13}$CO$^+$ in our model.     
Note that although there might be a hotter component in the nucleus 
(T$_k\sim$100K; see Mauersberger et al 1990), we must stress that n(H$_2$) inferred from the SiO line ratios depend weakly on
the adopted kinetic temperature if T$_k>$50K.

Figure 8 displays the range of LVG solutions that fit R$_{32}$ and R$_{54}$ within the errors. 
The assumed density n(H$_2$)=5$\times$10$^{5}$ fits satisfactorily 
R$_{32}$ and R$_{54}$, for a large interval of N(SiO)/$\Delta_v$.
Trying to delimit the LVG solution along the N/$\Delta_v$ axis is equivalent to 
choosing a beam filling factor ($\eta_{fill}$) for the clouds. Whereas it is justified to assume that the two lines 
share a common $\eta_{fill}$, its exact 
value in an external galaxy is unknown {\it a priori}. It is worth discussing to what extent 
our conclusions might depend on $\eta_{fill}$. 

We have run LVG models for n(H$_2$)=5$\times$10$^{5}$cm$^{-3}$ and T$_k$=50K, fitting 
R$_{Tmb}$=T$_{mb}$(SiO)/T$_{mb}$(H$^{13}$CO$^+$) in the different regions of the CND. 
Table 1 fully explores the dependence of N(SiO)/N(H$^{13}$CO$^+$) 
on $\eta_{fill}$; the filling factor is varied within its whole range (0.005--1). 
The main conclusion, independent of $\eta_{fill}$, 
is that the SiO abundance is significantly enhanced in the outer CND (region {\bf II}), 
where we derive N(SiO)/N(H$^{13}$CO$^+$)=5--15, and contrary to what it might be expected, it is lower at the center and 
at the starburst ring {\bf I}, where N(SiO)/N(H$^{13}$CO$^+$)=1--2. The SiO abundance rises in the outflow, where we 
infer N(SiO)/N(H$^{13}$CO$^+$)=3--5. 

\begin{figure}[tphb]
\caption{This figure represents LVG solutions fitting the R$_{54}$=I(SiO(5-4))/I(SiO(3-2)) and  
R$_{32}$=I(SiO(3-2))/I(SiO(2-1)) line ratios (continuous lines) observed in the SiO clouds of the nucleus of
NGC\,253, assuming T=50K. We consider errors of 20$\%$ and 30$\%$ in ratios R$_{54}$ and R$_{32}$ 
(pointed and dashed lines delimit the region of allowed solutions within the errors tolerance).
 An average density of n(H$_2$)=5$\times$10$^{5}$cm$^{-3}$ is a good compromise, indicating that the SiO emitting gas 
is dense. Trying to delimit further the LVG solution along the N/$\Delta_v$ axis is equivalent to
choose a beam filling factor ($\eta_{fill}$) for the clouds.}
\end{figure}

A density fall-off with radius has been reported by Wall et al 1991 
(and confirmed by Harrison et al 1999). 
They concluded that n(H$_2$) might reach $\sim$5$\times$10$^{4}$cm$^{-3}$ for r$>$20$\arcsec$, not far from region {\bf II}. 
If we take n(H$_2$)$\sim$5$\times$10$^{4}$cm$^{-3}$, we find that N(SiO)/N(H$^{13}$CO$^+$)=20-100 for {\bf II}.
Opacity in the 2--1 line of SiO is close to 
$\sim$1 for this solution, whereas $\tau\sim$0.02 for H$^{13}$CO$^+$. To fit the ratio of line temperatures we need 
imperatively to rise N(SiO)/N(H$^{13}$CO$^+$) in region {\bf II}.
Consequently, conclusions on the SiO abundance enhancement 
of region {\bf II} remain unchanged, if not clearly favoured,
in the lower density scenario.

From the ratio N(SiO)/N(H$^{13}$CO$^+$), we can infer the absolute abundance of SiO (X(SiO)=N(SiO)/N(H$_2$)), if we adopt 
a canonical value for H$^{13}$CO$^+$ ($<$X(H$^{13}$CO$^+$)$>$). 
It is expected that SiO will be more sensitive than H$^{13}$CO$^+$ 
to any variation of the chemical environment in NGC\,253. This hypothesis is 
successfully confronted to the measurements of abundances of H$^{12}$CO$^+$ obtained in a large variety of molecular 
clouds in our Galaxy where values close to 10$^{-8}$ are obtained in all cases (from which we derive 
X(H$^{13}$CO$^+$)$\sim$10$^{-10}$ if we adopt $^{13}$C/$^{12}$C$\sim$1/90). 
Though there is observational and theoretical evidence that X(H$^{12}$CO$^+$) may be reduced to 10$^{-10}$ in the 
hot cores of GMCs, SiO is lower by a similar factor in these sources (see Blake et al 1987); moreover, very compact sources are 
not expected to dominate the emissivity of H$^{13}$CO$^+$(1--0) at the scale we are observing 
the nucleus of NGC\,253. 
Contrary to X(H$^{13}$CO$^+$), X(SiO) is measured to vary by several orders of magnitude in the Galaxy, 
depending on the type and the location of molecular clouds in the disk (see discussion).
We can estimate  $<$X(H$^{13}$CO$^+$)$>$ in the nucleus of NGC\,253. From our data we obtain a column density 
of $<$N(H$^{13}$CO$^+$)$>$=3$\times$10$^{12}$, averaged within the CND. Similarly, we can 
calculate $<$N(H$_2$)$>$ from the $^{12}$CO(1--0) map of Canzian et al (1988), averaged on the same area. 
If we take the conversion factor x=N(H$_2$)/I$_{CO}$=3$\times$10$^{20}$cm$^{2}$Kkm$^{-1}$s, 
it follows $<$N(H$_2$)$>\sim$3$\times$10$^{22}$, which implies 
global abundances of $<$X(H$^{13}$CO$^+$)$>\sim$10$^{-10}$ in NGC\,253, quite similar to the value inferred for our Galaxy.

\begin{table*}[tphb]
\caption{Table 1 fully explores the dependence of N(SiO)/N(H$^{13}$CO$^+$) and X(SiO)
on $\eta_{fill}$; the filling factor is varied within its whole range (0.005--1) and the results are displayed
for the different regions of the nucleus where distinct values of R$_{I}$ are measured. See text for details.}
\end{table*}

Column 5 of Table 1 illustrates the variation of X(SiO) in the CND of NGC\,253. 
We derive X(SiO)$\sim$1--2$\times$10$^{-10}$ for the center and the starburst ring {\bf I}, 
X(SiO)$\sim$3--15$\times$10$^{-10}$ for region {\bf II}, and X(SiO)$\sim$3-5$\times$10$^{-10}$ for the outflow.       
The global abundance of SiO in the CND would be $<$X(SiO)$>\sim$1.5$\times$10$^{-10}$, namely, an order of magnitude above
the typical value for a PDR (Janssen et al 1995, Walmsley et al 1999). 
Most notably, X(SiO) reaches the highest abundance in the outer region {\bf II}, where the chemical processing of the gas 
by the nuclear starburst, mostly adscribed to region {\bf I}, should be minor. The SiO abundance is also significantly 
enhanced in the outflow, certainly powered by the starburst, but not directly on it. 
This result emphasizes that SiO emission is not always directly related to the dense gas near sites of recent star formation. 
Therefore, different mechanisms must be explored to account for the different chemical processing of molecular gas 
within the nucleus.

\section{Discussion}

We have established the existence of three distinct regions in the nuclear region of NGC\,253, based on morphological, 
kinematical and chemical criteria. Region {\bf I} dominates the global SiO emission and it is associated with the 
nuclear starburst taking place across $\sim$150pc. It presents a ring-like pattern,
interpreted as the gas response near the iILR of the bar. The SiO abundance in the starburst (X(SiO)$\sim$1--2$\times$10$^{-10}$) 
is significantly larger than that estimated for a typical PDR (X(SiO-PDR)$\sim$10$^{-11}$).
Region {\bf II} extending up to the map edges (r$\sim$300pc(20$\arcsec$)), displays a spiral-like morphology also 
present in other high-density tracers. The observed distribution and kinematics of molecular gas are interpreted as 
the signature of the bar oILR. The orbit crowding across the oILR coincides with the detection of strong 
non-circular motions ($\Delta$v$\sim$50--100kms$^{-1}$) and, most notably, with an enhancement in the SiO 
abundance (X(SiO)$\sim$a few 10$^{-9}$).  
The third region constitutes the molecular counterpart of the nuclear gas outflow observed 
in X-ray and optical lines. This was recently identified as a dust chimney. Two filaments (N and S) come out of the plane 
of the NGC\,253 nucleus from r$\pm$60pc, near a working surface where the gas is entrained by the outflow. Again, we have 
found here a link between the existence of high-velocities ($\Delta$v$\sim$100--150kms$^{-1}$) and an SiO enrichment 
(X(SiO)$>$3--5$\times$10$^{-10}$).

Two processes are thought to dominate the chemistry of the molecular material associated with a starburst:  
the photodissociation of the gas by intense UV field produced by OB associations, and the existence of strong shock waves.
Shocks can be generated locally during mass loss episodes in young stars (often identified as bipolar outflows). 
Alternatively, shocks may take place on a larger scale, either produced by supernova explosions or 
related to spiral/barred density waves.  
Contrary to X(H$^{13}$CO$^+$), X(SiO) is measured to vary by several orders of magnitude in the Galaxy, 
depending on the type and the location of molecular clouds in the disk: 
star forming clouds (X(SiO)$\sim$10$^{-7}$--10$^{-8}$),  
galactic center clouds apparently not forming stars (X(SiO)$\sim$10$^{-9}$) or quiescent clouds (X(SiO)$<$10$^{-12}$).

It is well established that shocks occurring in bipolar outflows can enhance the abundance of SiO in the gas phase 
to reach, in the most extreme cases, $\sim$10$^{-7}$ (Mart\'{\i}n-Pintado et al 1992). 
Although the starburst region {\bf I} might be described as a giant PDR (Carral et al 1994), 
the relatively large abundances of SiO measured  and the high densities of 
SiO clouds indicate that most of the emission arises in bipolar outflows powered by young 
massive stars, similarly to what is observed in the galactic complexes W51 and W49. 
Moreover, the particularly large ratio of I([SiII])/I([OI])$\sim$1 measured in the inner region of NGC\,253 
indicates an enhancement of Si in gas phase. This was interpreted by Carral et al 1994 as evidence of grain 
destruction in the starburst region, most likely by shocks related to massive star formation.

The case for shock processing is even stronger in region {\bf II} and in the outflow where 
X(SiO)$\sim$(50--100)$\times$X(SiO-PDR).
However, the origin of the large abundance of SiO in region {\bf II} is less clear. SiO emission could arise partly in 
bipolar outflows powered by young massive stars like in region {\bf I}. The major drawback to this explanation is
that region {\bf II} is located away from the nuclear starburst. Therefore one should not expect an enhancement of 
SiO when the chemical processing caused by star formation has declined. Alternatively, the SiO enhancement could be 
produced by large-scale shocks in the molecular gas, caused by the crossing of clouds orbits near the oILR 
of the bar.       
The only evidence for chemical enrichment of gas attributed to large-scale shocks is
found in the Galactic Center where values of $\sim$10$^{-9}$ have been measured in SiO clouds 
(Mart\'{\i}n-Pintado et al 1997, H{\"u}ttemeister et al 1998).
However there is no theoretical consensus whether the shocks can be caused by the action of density waves, 
especially in molecular clouds. The hierarchy of fragmentation inhibits shocks in molecular clouds; 
the mean-free paths for cloud-cloud collisions are of the same order of 
magnitude as the thickness of the disturbing potential well (Casoli and Combes 1982, Combes and Gerin 1985).

If the starburst scenario gives rise to SiO emission in bipolar outflows of young massive stars, the differences
in the abundance of SiO in regions {\bf I} and {\bf II} might be explained if the starbursts are in a markedly different stage
of evolution. The older starburst would occur in region {\bf I}, where a large fraction of the young massive stars are 
already on the main sequence. The latter explains why the classical tracers of the starburst phenomenon are mostly 
restricted to {\bf I}. In contrast, region {\bf II} would contain a younger and less-evolved starburst, where most of the young 
massive stars would be still in the mass loss phase, characteristic of the pre-main sequence stage, where
energetic bipolar outflows occur.

Finally, the interpretation of the morphology and kinematics of the molecular gas filaments 
give additional support to the outflow scenario. The enhancement of SiO in the gas phase observed in this region 
can be explained by the chemical processing of grains by the strong high-velocity shocks (v$>$200kms$^{-1}$) 
associated to the outflow.

\acknowledgements

This work has been partially supported by the Spanish CICYT under grant number 
PB96-0104.

\end{document}